\documentstyle[11pt,aaspp4]{article}

\slugcomment{\hspace*{-1cm}To appear in the Astrophysical Journal (accepted
1998 November 20)}

\lefthead{MIGHELL}
\righthead{PARAMETER ESTIMATION IN ASTRONOMY WITH POISSON-DISTRIBUTED DATA}

\begin{document}

\def\et{\hbox{et~al.\ }}
\def\lea{\mathrel{<\kern-1.0em\lower0.9ex\hbox{$\sim$}}}
\def\gea{\mathrel{>\kern-1.0em\lower0.9ex\hbox{$\sim$}}}
\newcommand{\avec}{{\mbox{\boldmath{$a$}}}}
\def\ifundefined#1{\expandafter\ifx\csname#1\endcsname\relax}
\def\firstuse#1{}
\def\noteforeditor#1{}

\newenvironment{outline}{
\renewcommand{\baselinestretch}{0.8}
\begin{itemize}
\Large
}{
\end{itemize}
}

\def\change{{\hspace*{-2mm}\raisebox{-0.6em}{$\spadesuit$}}}

\title{PARAMETER ESTIMATION IN ASTRONOMY WITH POISSON-DISTRIBUTED DATA.
I. THE {\Large{$\chi^2_\gamma$}} STATISTIC}

\author{
\sc
Kenneth J. Mighell
}
\affil{
\small
Kitt Peak National Observatory,
National Optical Astronomy Observatories\altaffilmark{1},\\
P. O. Box 26732, Tucson, AZ~~85726-6732\\
Electronic mail:
mighell@noao.edu
}

\altaffiltext{1}{NOAO is operated
by the Association of Universities for Research in Astronomy, Inc., under
cooperative agreement with the National Science Foundation.}

\begin{abstract}
Applying the standard weighted mean formula,
$
\left[\sum_i {n_i \sigma^{-2}_i}\right]
/
\left[\sum_i {\sigma^{-2}_i}\right]
$,
to determine the weighted mean
of data, $n_i$, drawn from a Poisson
distribution, will, on average,
underestimate the true mean by $\sim$$1$ for all true mean
values larger than $\sim$$3$
when the common assumption is made
that the error of the $i$th observation is
$\sigma_i = \max(\sqrt{n_i},1)$.
This small, but statistically significant offset,
explains the long-known observation that chi-square minimization techniques
which use the modified Neyman's $\chi^2$ statistic,
$\chi^2_{\rm{N}} \equiv \sum_i (n_i-y_i)^2/\max(n_i,1)$,
to compare Poisson-distributed data with model values, $y_i$, will
typically predict a total number of counts that
underestimates the true total
by about $1$ count per bin.
Based on
my
finding that the weighted mean of data
drawn from a Poisson distribution can be
determined using the formula
$
\left[
\sum_i \left[n_i+\min\left(n_i,1\right)\right]\left(n_i+1\right)^{-1}
\right]
/
\left[
\sum_i \left(n_i+1\right)^{-1}
\right]
$, I
propose that a new $\chi^2$ statistic,
$\chi^2_\gamma
\equiv
\sum_i
\left[ n_i + \min\left( n_i, 1\right) - y_i\right]^2
/
\left[ n_i + 1 \right]$,
should always be used to analyze Poisson-distributed data
in preference to the modified Neyman's $\chi^2$ statistic.
I
demonstrate the power and usefulness of $\chi^2_\gamma$ minimization
by using two statistical fitting techniques and five $\chi^2$ statistics
to analyze simulated X-ray power-law 15-channel spectra
with large and small counts per bin.
I
show that $\chi^2_\gamma$ minimization with
the Levenberg-Marquardt or Powell's method can produce
excellent results (mean slope errors $\lea$$3$\%)
with spectra having as few as 25 total counts.
\end{abstract}

\newpage
\section{INTRODUCTION}

The determination of the weighted mean is the fundamental problem
for chi-square ($\chi^2$) minimization methods.
The goodness-of-fit between an observation of $N$ data values, $x_i$,
with errors, $\sigma_i$, and a model, $m_i$,
can be determined by using the standard chi-square statistic:
\begin{equation}
\chi^2
\equiv
{\sum_{i=1}^{N}}
\left[
\frac{ x_i - m_i }{ \sigma_i }
\right]^2\,.
\label{eq:chisq_std}
\end{equation}
The theory of least-squares states that the optimum value of
all the parameters of the model are
obtained when the chi-square statistic
is minimized with respect to each parameter simultaneously.
For example,
the standard formula of the weighted mean
can be derived by
assuming that the model is a constant
and then solving the equation,
\begin{equation}
\frac{\partial}{\partial\mu_{\rm{w}}}
{\sum_{i=1}^{N}}
\left[
\frac{x_i - \mu_{\rm{w}}}{\sigma_i}
\right]^2
=
0
\ ,
\end{equation}
for that constant:
\begin{equation}
\mu_{\rm{w}}
\equiv
\frac{\displaystyle
\sum_{i=1}^N
\frac{x_i}{\sigma_i^2}
}{\displaystyle
\sum_{i=1}^N
\frac{1}{\sigma_i^2}
}
\,.
\label{eq:weighted_mean}
\end{equation}
The standard weighted-mean formula
thus weights every data value, $x_i$, inversely by its own
variance (i.e.\ $\sigma_i^2$).

Let us assume that all the data values come from a pure counting
experiment where each data value, $n_i$, is a random integer deviate
drawn from a Poisson (\cite{poisson_1837}) distribution,
\begin{equation}
P(k;\mu)
\equiv
\frac{\mu^k}{k!}e^{-\mu}
\label{eq:poisson}
,
\end{equation}
with a mean value of $\mu$.
Let us also make the common assumption that the error of each data
value is the square root of the mean of the parent Poisson distribution.
Using these transformations,
$x_i \Rightarrow n_i$
and
$\sigma_i \Rightarrow \sqrt{\mu}\,$,
we see that Equation (\ref{eq:weighted_mean}) becomes
\begin{equation}
\mu_{\rm{P}}
\equiv
\frac{\displaystyle
\sum_{i=1}^N
\frac{n_i}{\mu}
}{\displaystyle
\sum_{i=1}^N
\frac{1}{\mu}
}
\ ,
\label{eq:weighted_mean_pearson}
\end{equation}
which reduces to become the definition of the sample mean:
\begin{equation}
\mu_{\rm{P}}
\equiv
\frac{1}{N}
\sum_{i=1}^N
n_i
\label{eq:standard_mean}
\ .
\end{equation}
In the limit of a large number of observations
of the Poisson distribution $P(k;\mu)$,
we find that
Equation (\ref{eq:standard_mean})
will, on average,
determine the mean of the parent Poisson distribution
for all true mean values $\mu$:
\begin{eqnarray}
\lim_{N \rightarrow \infty}
\left[
\mu_{\rm{P}}
\right]
&\equiv
&
\lim_{N \rightarrow \infty}
\left[
\frac{1}{N}
\sum_{i=1}^N
n_i
\right]
\nonumber
\\
&\approx
&
\lim_{N \rightarrow \infty}
\left[
\frac{1}{N}
\sum_{k=0}^{\infty}
k
\left\{
\rule[-0.1em]{0em}{1.2em}
N P(k;\mu)
\right\}
\right]
\nonumber
\\
&=
&
\sum_{k=0}^{\infty}
k
\left\{
\rule[-0.1em]{0em}{1.2em}
P(k;\mu)
\right\}
\nonumber
\\
&=
&
\sum_{k=0}^{\infty}
k
\frac{\mu^k}{k!}e^{-\mu}
\nonumber
\\
&=
&
0
\frac{\mu^0}{0!}e^{-\mu}
+
e^{-\mu}
\sum_{k=1}^{\infty}
\frac{\mu^k}{(k-1)!}
\nonumber
\\
&=
&
e^{-\mu}
\mu
\sum_{k=1}^{\infty}
\frac{\mu^{k-1}}{(k-1)!}
\nonumber
\\
&=
&
e^{-\mu}
\mu
\sum_{j=0}^{\infty}
\frac{\mu^j}{j!}
\nonumber
\\
&=
&
e^{-\mu}
\mu
e^{\mu}
\nonumber
\\
&=
&\mu
\label{eq:weighted_mean_pearson_limit}
\ .
\end{eqnarray}

{\em{
Applying the standard weighted mean formula,
$
\left[\sum_i {n_i \sigma^{-2}_i}\right]
/
\left[\sum_i {\sigma^{-2}_i}\right]
$,
to determine the weighted mean
of data, $n_i$, drawn from a Poisson distribution, will,
on average,
determine the mean of the parent Poisson distribution
for all true mean values
if a constant weight is assigned to all data values}}
(i.e. $\sigma^{-2}\equiv$ constant).

It is a common practice to
assume that the error of a Poisson deviate $n$ is
$\sigma \equiv \sqrt{n}$.
Unfortunately, this practice
causes the standard weighted-mean formula
to be undefined for data values of zero.
A simple solution to this computational problem is to
arbitrarily assign a non-zero constant error to all Poisson
deviates with a value of zero.
Let us make the common assumption that the error of each data value, $n_i$,
is equal to $\sqrt{n_i}$ or 1 --- whichever is greater.
Using the following transformations,
$x_i \Rightarrow n_i\,$
and
$\sigma_i \Rightarrow \max(\sqrt{n_i},1)\,$,
we see that Equation (\ref{eq:weighted_mean}) becomes
\begin{equation}
\mu_{\rm{N}}
\equiv
\frac{\displaystyle
\sum_{i=1}^N
\frac{n_i}{\max(n_i,1)}
}{\displaystyle
\sum_{i=1}^N
\frac{1}{\max(n_i,1)}
}
\ .
\label{eq:weighted_mean_modified_neyman}
\end{equation}
In the limit of a large number of observations
of the Poisson distribution $P(k;\mu)$,
we find that
\begin{eqnarray}
\lim_{N \rightarrow \infty}
\left[
\mu_{\rm{N}}
\right]
&\equiv
&
\lim_{N \rightarrow \infty}
\left[
\frac{\displaystyle
\sum_{i=1}^N
\frac{n_i}{\max(n_i,1)}
}{\displaystyle
\sum_{i=1}^N
\frac{1}{\max(n_i,1)}
}
\right]
\nonumber
\\
&\approx
&
\lim_{N \rightarrow \infty}
\left[
\frac{\displaystyle
\sum_{k=0}^\infty
\frac{k}{\max(k,1)}
\left\{
\rule[-0.1em]{0em}{1.2em}
N P(k;\mu)
\right\}
}{\displaystyle
\sum_{k=0}^\infty
\frac{1}{\max(k,1)}
\left\{
\rule[-0.1em]{0em}{1.2em}
N P(k;\mu)
\right\}
}
\right]
\nonumber
\\
&=
&
\frac{\displaystyle
\sum_{k=0}^\infty
\frac{k}{\max(k,1)}
\left\{
\rule[-0.1em]{0em}{1.2em}
P(k;\mu)
\right\}
}{\displaystyle
\sum_{k=0}^\infty
\frac{1}{\max(k,1)}
\left\{
\rule[-0.1em]{0em}{1.2em}
P(k;\mu)
\right\}
}
\nonumber
\\
&=&
\frac{\displaystyle
\frac{0}{1}
P(0;\mu)
+
\sum_{k=1}^\infty
\frac{k}{k}
P(k;\mu)
}{\displaystyle
\frac{1}{1}
P(0;\mu)
+
\sum_{k=1}^\infty
\frac{1}{k}
P(k;\mu)
}
\nonumber
\\
&=&
\frac{\displaystyle
1-e^{-\mu}
}{\displaystyle
e^{-\mu}
+
e^{-\mu}
\sum_{k=1}^\infty
\frac{\mu^k}{k\,k!}
}
\nonumber
\\
&=
&
\frac{\displaystyle
e^\mu - 1
}{\displaystyle
1
+
\left[
\sum_{k=1}^\infty
\frac{\mu^k}{k\,k!}
\right]
}
\label{eq:wmean_limit_neyman_with_summation}
\\
&=
&
\frac{\displaystyle
e^\mu - 1
}{\displaystyle
1
+
\left[
\rule[-0.1em]{0em}{1.2em}
{\rm{Ei}}(\mu) - \gamma - \ln(\mu)
\right]~,
}
\label{eq:wmean_limit_neyman}
\end{eqnarray}
where
Ei$(x)$ is the exponential integral of $x$,
$
\mbox{Ei}(x)
=
-\int_{-x}^{\infty} \frac{e^{-t}}{t}\,dt
=
\int_{-\infty}^{x} \frac{e^{-t}}{t}\,dt
$
(for $x > 0$), and
$\gamma$ is the Euler-Mascheroni constant:
$\gamma
\equiv
\lim_{n \rightarrow \infty}
\left[
\left\{
\sum_{i=1}^{n}
\frac{1}{n}
\right\}
- \ln(n)
\right]
=
0.5772156649\cdots
$
(see, e.g., Abramowitz \& Stegun \cite{abst1964}).

Let us now investigate the limit of Equation
(\ref{eq:wmean_limit_neyman})
with large Poisson mean values.
The transformation of
Equation (\ref{eq:wmean_limit_neyman_with_summation})
to
Equation (\ref{eq:wmean_limit_neyman})
used the power series of Ei$(x)$,
\begin{equation}
\mbox{Ei}(x)
=
  \gamma
+ \ln(x)
+ \frac{x}{1\cdot1!}
+ \frac{x^2}{2\cdot2!}
+ \frac{x^3}{3\cdot3!}
+ \cdots~,
\end{equation}
which has the following asymptotic expansion:
\begin{equation}
\mbox{Ei}(x)
\approx
\frac{e^x}{x}
\left(
  1
+ \frac{1!}{x}
+ \frac{2!}{x^2}
+ \frac{3!}{x^3}
+ \cdots
\right)~.
\end{equation}
{}From the following limit,
\begin{equation}
\lim_{x\rightarrow\infty}
\left[
x
\left(
  1
+ \frac{1!}{x}
+ \frac{2!}{x^2}
+ \frac{3!}{x^3}
+ \cdots
\right)^{\!\!-1}
\!\!
-~x
\right]
=
-1~,
\end{equation}
we see that Ei$(x)$ asymptotically approaches the function $e^{x}/(x-1)$
for large values of $x$.
For $x\geq13$ this approximation has an error of $<$1\%\,;
for $x\geq33$ the error is $\leq$0.1\%\,.
In the limit of large mean Poisson values,
we see that the numerator
of Equation
(\ref{eq:wmean_limit_neyman})
is dominated by the $e^\mu$ term
while the denominator is dominated by the Ei$(\mu)$ term
which asymptotically approaches the value of $e^\mu/(\mu-1)$.
{\em{We then have come to the surprising conclusion that
for Poisson distributions with large mean values,
$
\lim_{N \rightarrow \infty}
\left[
\mu_{\rm{N}}
\right]
$
approaches the value of $\mu-1$ instead of the expected value of $\mu$.}}

Equation (\ref{eq:wmean_limit_neyman})
can also be investigated graphically.
Figure \ref{fig-weighted_mean_modified_neyman}a
\firstuse{Fig\ref{fig-weighted_mean_modified_neyman}}
plots the difference between the weighted mean computed using
Equation (\protect\ref{eq:weighted_mean_modified_neyman})
and the true mean for Poisson-distributed data with
true mean values between 0.001 and 1000.
Each open square represents the weighted mean of $4$$\times$$10^6$
Poisson deviates at each given true mean value.
The solid curve through the data
[open squares in Fig.\ \ref{fig-weighted_mean_modified_neyman}a]
is the difference between
Equation (\protect\ref{eq:wmean_limit_neyman}) and the true mean.
Note that
Equation (\protect\ref{eq:wmean_limit_neyman})
underestimates the true mean by $\sim$$1$ for large true mean values
(as predicted above).

{\em{
Applying the standard weighted mean formula,
$
\left[\sum_i {n_i \sigma^{-2}_i}\right]
/
\left[\sum_i {\sigma^{-2}_i}\right]
$,
to determine the weighted mean
of data, $n_i$, drawn from a Poisson distribution, will,
on average,
underestimate the true mean by $\sim$$1$ for all true mean values
larger than $\sim$$3$ when the common assumption is made
that the error of the $i$th observation is
$\sigma_i = \max(\sqrt{n_i},1)$.
}}

\section{THE WEIGHTED MEAN OF POISSON-DISTRIBUTED DATA}

We will now develop a weighted-mean formula for Poisson-distributed
data that will, on average,
determine the true mean of the parent distribution
for all true mean values.

Let us assume that the error of each data value,
$n_i$,
is equal to
$\sqrt{n_i+1}$
instead of
$\max(\sqrt{n_i},1)\,$.
Using the following transformations,
$x_i \Rightarrow n_i$
and
$\sigma_i \Rightarrow \sqrt{n_i+1}\,$,
we see that Equation (\ref{eq:weighted_mean}) becomes
\begin{equation}
\mu_\alpha
\equiv
\frac{\displaystyle
\sum_{i=1}^N
\frac{n_i}{n_i+1}
}{\displaystyle
\sum_{i=1}^N
\frac{1}{n_i+1}
}
\ .
\label{eq:weighted_mean_alpha}
\end{equation}
In the limit of a large number of observations
of the Poisson distribution $P(k;\mu)$,
we find that
\begin{eqnarray}
\lim_{N \rightarrow \infty}
\left[
\mu_\alpha
\right]
&\equiv&
\lim_{N \rightarrow \infty}
\left[
\frac{\displaystyle
\sum_{i=1}^N
\frac{n_i}{n_i+1}
}{\displaystyle
\sum_{i=1}^N
\frac{1}{n_i+1}
}
\right]
\nonumber
\\
&\approx&
\lim_{N \rightarrow \infty}
\left[
\frac{\displaystyle
\sum_{k=0}^\infty
\frac{k}{k+1}
\left\{
\rule[-0.1em]{0em}{1.2em}
N P(k;\mu)
\right\}
}{\displaystyle
\sum_{k=0}^\infty
\frac{1}{k+1}
\left\{
\rule[-0.1em]{0em}{1.2em}
N P(k;\mu)
\right\}
}
\right]
\nonumber
\\
&=&
\frac{\displaystyle
\sum_{k=0}^\infty
\frac{k}{k+1}
\left\{
\rule[-0.1em]{0em}{1.2em}
P(k;\mu)
\right\}
}{\displaystyle
\sum_{k=0}^\infty
\frac{1}{k+1}
\left\{
\rule[-0.1em]{0em}{1.2em}
P(k;\mu)
\right\}
}
\nonumber
\\
&=&
\frac{\displaystyle
\frac{1}{\mu}\left( \mu - 1 + e^{-\mu} \right)
}{\displaystyle
\frac{1}{\mu}\left( 1 - e^{-\mu} \right)
}
\nonumber
\\
&=&
\frac{\mu}{1-e^{-\mu}} - 1
\ .
\label{eq:weighted_mean_limit_alpha}
\end{eqnarray}
Figure
\ref{fig-weighted_mean_modified_neyman}b
graphically confirms this finding.
Increasing the error estimates from
$\max(\sqrt{n_i},1)$ to $\sqrt{n_i+1}$ has only yielded a minor improvement.
Notice that the dip in the solid curve in Fig.\
\ref{fig-weighted_mean_modified_neyman}a
at $\mu \approx 6$ is not present
in the solid curve in Fig.\
\ref{fig-weighted_mean_modified_neyman}b.
A more radical change appears to be required
in order for us to develop a weighted-mean formula
for Poisson-distributed data.

Let us now add one to all data values
and assume that the error of each data value is the square root
of the new data value.
Using these transformations,
$x_i \Rightarrow n_i+1$
and
$\sigma_i \Rightarrow \sqrt{n_i+1}\,$,
we see that Equation (\ref{eq:weighted_mean}) becomes
\begin{eqnarray}
\mu_\beta
&\equiv&
\frac{\displaystyle
\sum_{i=1}^N
\frac{n_i+1}{n_i+1}
}{\displaystyle
\sum_{i=1}^N
\frac{1}{n_i+1}
}
{}~.
\label{eq:weighted_mean_beta}
\end{eqnarray}
In the limit of a large number of observations
of the Poisson distribution $P(k;\mu)$,
we find that
\begin{eqnarray}
\lim_{N \rightarrow \infty}
\left[
\mu_\beta
\right]
&\equiv&
\lim_{N \rightarrow \infty}
\left[
\frac{\displaystyle
\sum_{i=1}^N
\frac{n_i+1}{n_i+1}
}{\displaystyle
\sum_{i=1}^N
\frac{1}{n_i+1}
}
\right]
\nonumber
\\
&\approx&
\lim_{N \rightarrow \infty}
\left[
\frac{\displaystyle
\sum_{k=0}^\infty
\frac{k+1}{k+1}
\left\{
\rule[-0.1em]{0em}{1.2em}
N P(k;\mu)
\right\}
}{\displaystyle
\sum_{k=0}^\infty
\frac{1}{k+1}
\left\{
\rule[-0.1em]{0em}{1.2em}
N P(k;\mu)
\right\}
}
\right]
\nonumber
\\
&=&
\frac{\displaystyle
\sum_{k=0}^\infty
\left\{
\rule[-0.1em]{0em}{1.2em}
P(k;\mu)
\right\}
}{\displaystyle
\sum_{k=0}^\infty
\frac{1}{k+1}
\left\{
\rule[-0.1em]{0em}{1.2em}
P(k;\mu)
\right\}
}
\nonumber
\\
&=&
\frac{\displaystyle
1
}{\displaystyle
\frac{1}{\mu}\left( 1 - e^{-\mu} \right)
}
\nonumber
\\
&=&
\frac{\mu}{1-e^{-\mu}}
\ .
\label{eq:weighted_mean_limit_beta}
\end{eqnarray}
Figure
\ref{fig-weighted_mean_modified_neyman}c
graphically confirms this finding.
We have now made significant progress towards our goal of
developing a weighted-mean formula for Poisson-distributed data.
Applying Equation
(\ref{eq:weighted_mean_beta})
to determine the weighted mean
of Poisson-distributed data, will,
on average,
estimate the true mean with $\lea$$1$\% errors
for true Poisson mean values $\mu \gea 5$.

The deviation of the solid curve in Figure
\ref{fig-weighted_mean_modified_neyman}c
from zero can be eliminated by making just a
minor change to our transformations.
Using the same errors as above,
$\sigma_i \Rightarrow \sqrt{n_i+1}\,$,
but now adding one to only those data values that are initially
greater than zero,
$n_i \Rightarrow n_i + \min(n_i,1)\,$,
we see that Equation (\ref{eq:weighted_mean}) becomes
\begin{equation}
\mu_\gamma
\equiv
\frac{\displaystyle
\sum_{i=1}^N
\frac{n_i+\min(n_i,1)}{n_i+1}
}{\displaystyle
\sum_{i=1}^N
\frac{1}{n_i+1}
}
\ .
\label{eq:weighted_mean_gamma}
\end{equation}
In the limit of a large number of observations
of the Poisson distribution $P(k;\mu)$,
we find that
\begin{eqnarray}
\lim_{N \rightarrow \infty}
\left[
\mu_\gamma
\right]
&\equiv&
\lim_{N \rightarrow \infty}
\left[
\frac{\displaystyle
\sum_{i=1}^N
\frac{n_i+\min(n_i,1)}{n_i+1}
}{\displaystyle
\sum_{i=1}^N
\frac{1}{n_i+1}
}
\right]
\nonumber
\\
&\approx&
\lim_{N \rightarrow \infty}
\left[
\frac{\displaystyle
\sum_{k=0}^\infty
\frac{k+\min(k,1)}{k+1}
\left\{
\rule[-0.1em]{0em}{1.2em}
N P(k;\mu)
\right\}
}{\displaystyle
\sum_{k=0}^\infty
\frac{1}{k+1}
\left\{
\rule[-0.1em]{0em}{1.2em}
N P(k;\mu)
\right\}
}
\right]
\nonumber
\\
&=&
\frac{\displaystyle
\sum_{k=0}^\infty
\frac{k+\min(k,1)}{k+1}
\left\{
\rule[-0.1em]{0em}{1.2em}
P(k;\mu)
\right\}
}{\displaystyle
\sum_{k=0}^\infty
\frac{1}{k+1}
\left\{
\rule[-0.1em]{0em}{1.2em}
P(k;\mu)
\right\}
}
\nonumber
\\
&=&
\frac{\displaystyle
\frac{0}{1}P(0;\mu)
+
\sum_{k=1}^\infty
\frac{k+1}{k+1}
P(k;\mu)
}{\displaystyle
\sum_{k=0}^\infty
\frac{1}{k+1}
P(k;\mu)
}
\nonumber
\\
&=&
\frac{\displaystyle
1 - e^{-\mu}
}{\displaystyle
\frac{1}{\mu}\left( 1 - e^{-\mu} \right)
}
\nonumber
\\
&=&
\mu
\ .
\label{eq:weighted_mean_limit_gamma}
\end{eqnarray}
Figure
\ref{fig-weighted_mean_modified_neyman}d
graphically confirms this finding.
We have now achieved our goal of
developing a weighted-mean formula for Poisson-distributed data.
{\em{Applying Equation
(\ref{eq:weighted_mean_gamma})
to determine the weighted mean
of Poisson-distributed data, will,
on average,
estimate the true mean for all true Poisson mean values
}} ($\mu \geq 0$).

\section{THE $\chi^2_\gamma$ STATISTIC}

Based on my
finding that the weighted mean of data
drawn from a Poisson distribution can be
determined using the formula
$
\left[
\sum_i \left[n_i+\min\left(n_i,1\right)\right]\left(n_i+1\right)^{-1}
\right]
/
\left[
\sum_i \left(n_i+1\right)^{-1}
\right]
$,
I
propose that,
given $N$ observations ($n_i$) and a model ($m_i$),
a new $\chi^2$ statistic,
\begin{equation}
\chi^2_\gamma
\equiv
{\sum_{i=1}^{N}}
\frac{ \left[ n_i + \min\left( n_i, 1\right) - m_i \right]^2 }{ n_i+1 }
\ ,
\end{equation}
should always be used to analyze Poisson-distributed data
in preference to the modified Neyman's $\chi^2$ statistic,
\begin{equation}
\chi_{\rm{N}}^2
\equiv
{\sum_{i=1}^{N}}
\frac{ (n_i - m_i)^2 }{ \max(n_i,1) }
\ ,
\end{equation}
because the weighted-mean formula for the modified
Neyman's
$\chi^2$ statistic
[$\mu_{\rm{N}}$:
Equation (\ref{eq:weighted_mean_modified_neyman})\,]
systematically underestimates the true mean value of Poisson-distributed
data with true mean values $\mu\gea0.5$
(see Fig.\ \ref{fig-weighted_mean_modified_neyman}a).

For Poisson-distributed data, it has long been observed that,
in many cases,
chi-square fits using the modified Neyman's $\chi^2$ statistic
and the
Pearson's $\chi^2$ statistic,
\begin{equation}
\chi_{\rm{P}}^2
\equiv
{\sum_{i=1}^{N}}
\frac{ (n_i - m_i)^2 }{ m_i }
\ ,
\end{equation}
will
underestimate and overestimate the total
area, respectively,
while the usage of
the maximum likelihood ratio statistic for Poisson distributions,
\begin{equation}
\chi^2_\lambda
\equiv
2{\sum_{i=1}^{N}}
\left[
m_i - n_i + n_i\ln\left(\frac{n_i}{m_i}\right)
\right]
\label{eq:x2l}
\ ,
\end{equation}
preserves the total area
(e.g., Baker \& Cousins \cite{baco1984} and references therein).

It has been known for decades that
chi-square minimization techniques
using the modified Neyman's $\chi^2$ statistic
to analyze Poisson-distributed data
will typically predict a total number of counts (total area) that
underestimates the true total counts by about $1$ count per bin
(e.g.,
Bevington \cite{be1969},
Wheaton \et \cite{weet1995}).
The reason why this underestimation occurs is now obvious:
the application of the modified Neyman's $\chi^2$ statistic to
Poisson-distributed data causes the fitted model value at each bin,
$m_i$, to be, on average,
underestimated by $\sim$$1$ count for all true Poisson
model mean values $\gea$$3$\,.
The underestimation of the true mean by one count gives
a very large $20$\% error when the true mean of the data is 5
but only a $1$\% error when the true mean of the data is 100.
It would clearly be difficult to detect such a small systematic error
with {\em{small samples}} of Poisson-distributed
data with {\em{large true mean values}}.
Figure \ref{fig-weighted_mean_modified_neyman}a
shows that this underestimation is real and
is easily measurable with {\em{large samples}} of Poisson-distributed data.

The number of degrees of freedom, commonly represented with the
symbol $\nu$,
of a chi-square minimization problem
is the difference between the number of observations (sample size)
and the number of free parameters ($M$) of the model:
$\nu \equiv N - M$.

The reduced chi-square of the Pearson's $\chi^2$ statistic is, by
definition, the value of Pearson's $\chi^2$ statistic divided by
the number of degrees of freedom:
\begin{equation}
\frac{
\chi_{\rm{P}}^2
}{
\nu
}
\equiv
\frac{1}{N-M}
{\sum_{i=1}^{N}}
\frac{ (n_i - m_i)^2 }{ m_i }
\ .
\label{eq:chi_nu_P}
\end{equation}
On average, the expected reduced chi-square value of
a proper $\chi^2$ statistic with a perfect model is one ---
given a large number of observations.
Now let us assume that our data comes from a Poisson distribution with a
mean value of $\mu$.
In this case, the model $m_i$ will be a constant,
$\mu_{\rm{P}}$ [Equation (\ref{eq:weighted_mean_pearson})],
which will,
on average, have a value,
$\mu_{\rm{P}^\prime}$,
given by
Equation (\ref{eq:weighted_mean_pearson_limit})
in the limit of a large number of observations
(N.B. $\mu_{\rm{P}^\prime} \equiv \mu$).
The model is a constant and therefore there is
only one degree-of-freedom: $M=1$.
Given these assumptions, we find that,
in the limit of a large number of observations,
the reduced chi-square of the Pearson's $\chi^2$ statistic with
the model $\mu_P$ is
\begin{eqnarray}
\lim_{N \rightarrow \infty}
\left[
\frac{\chi^2_{\rm{P}}}{\nu}
\right]
&\equiv&
\lim_{N\rightarrow\infty}
\left[
\frac{1}{N-M}
{\sum_{i=1}^{N}}
\frac{ (n_i - m_i)^2 }{ m_i }
\right]
\nonumber
\\
&=&
\lim_{N\rightarrow\infty}
\left[
\frac{1}{N-1}
{\sum_{i=1}^{N}}
\frac{ (n_i - \mu_{\rm{P}})^2 }{ \mu_{\rm{P}} }
\right]
\nonumber
\\
&\approx&
\lim_{N\rightarrow\infty}
\left[
\frac{1}{N-1}
{\sum_{k=0}^{\infty}}
\frac{ (k - \mu_{\rm{P}^\prime})^2 }{ \mu_{\rm{P}^\prime} }
\left\{
\rule[-0.1em]{0em}{1.2em}
N P(k;\mu)
\right\}
\right]
\nonumber
\\
&=&
{\sum_{k=0}^{\infty}}
\frac{ (k - \mu_{\rm{P}^\prime})^2 }{ \mu_{\rm{P}^\prime} }
\left\{
\rule[-0.1em]{0em}{1.2em}
P(k;\mu)
\right\}
\nonumber
\\
&=&
{\sum_{k=0}^{\infty}}
\frac{ (k - \mu)^2 }{ \mu }
\left\{
\rule[-0.1em]{0em}{1.2em}
P(k;\mu)
\right\}
\nonumber
\\
&=&
\frac{1}{\mu}
\left(
\left[
{\sum_{k=0}^{\infty}}
k^2 P(k;\mu)
\right]
- 2\mu
\left[
{\sum_{k=0}^{\infty}}
k P(k;\mu)
\right]
+ \mu^2
\left[
{\sum_{k=0}^{\infty}}
P(k;\mu)
\right]
\,\right)
\nonumber
\\
&=&
\frac{1}{\mu}
\left(
\left[
\rule[-0.1em]{0em}{1.2em}
\mu^2 + \mu
\right]
- 2\mu
\left[
\rule[-0.1em]{0em}{1.2em}
\mu
\right]
+ \mu^2
\left[
\rule[-0.1em]{0em}{1.2em}
1
\right]
\,\right)
\nonumber
\\
&=&
1~.
\end{eqnarray}

The reduced chi-square of the modified Neyman's $\chi^2$ statistic is, by
definition, the value of the modified Neyman's $\chi^2$ statistic divided by
the number of degrees of freedom:
\begin{equation}
\frac{
\chi_{\rm{N}}^2
}{
\nu
}
\equiv
\frac{1}{N-M}
{\sum_{i=1}^{N}}
\frac{ (n_i - m_i)^2 }{ \max(n_i,1) }
\ .
\end{equation}
Now let us assume that our data comes from a Poisson distribution with a
mean value of $\mu$.
In this case, the model $m_i$ will be a constant,
$\mu_{\rm{N}}$ [Equation (\ref{eq:weighted_mean_modified_neyman})],
which will,
on average, have a value,
$\mu_{\rm{N}^\prime}$,
given by
Equation (\ref{eq:wmean_limit_neyman})
in the limit of a large number of observations.
Given these assumptions, we find that,
in the limit of a large number of observations,
the reduced chi-square of the modified Neyman's $\chi^2$ statistic with
the model $\mu_N$ is
\begin{eqnarray}
\lim_{N \rightarrow \infty}
\left[
\frac{\chi^2_{\rm{N}}}{\nu}
\right]
&\equiv&
\lim_{N \rightarrow \infty}
\left[
\frac{1}{N-M}
{\sum_{i=1}^{N}}
\frac{ (n_i - m_i)^2 }{ \max(n_i,1) }
\right]
\nonumber
\\
&=&
\lim_{N \rightarrow \infty}
\left[
\frac{1}{N-1}
{\sum_{i=1}^{N}}
\frac{ (n_i - \mu_{\rm{N}})^2 }{ \max(n_i,1) }
\right]
\nonumber
\\
&\approx&
\lim_{N \rightarrow \infty}
\left[
\frac{1}{N-1}
{\sum_{k=0}^{\infty}}
\frac{ (k - \mu_{\rm{N}^\prime})^2 }{ \max(k,1) }
\left\{
\rule[-0.1em]{0em}{1.2em}
N P(k;\mu)
\right\}
\right]
\nonumber
\\
&=&
{\sum_{k=0}^{\infty}}
\frac{ (k - \mu_{\rm{N}^\prime})^2 }{ \max(k,1) }
\left\{
\rule[-0.1em]{0em}{1.2em}
P(k;\mu)
\right\}
\nonumber
\\
&=&
\mu_{\rm{N}^\prime}^2
e^{-\mu}
+
{\sum_{k=1}^{\infty}}
\frac{ (k - \mu_{\rm{N}^\prime})^2 }{ k }
P(k;\mu)
\nonumber
\\
&=&
\mu_{\rm{N}^\prime}^2
e^{-\mu}
+
\left[
\rule[-0.1em]{0em}{1.2em}
{\sum_{k=1}^{\infty}}
k
P(k;\mu)
\right]
-2\mu_{\rm{N}^\prime}
\left[
\rule[-0.1em]{0em}{1.2em}
{\sum_{k=1}^{\infty}}
P(k;\mu)
\right]
+\mu_{\rm{N}^\prime}^2
\left\{
\rule[-0.1em]{0em}{1.2em}
{\sum_{k=1}^{\infty}}
\frac{1}{k}
P(k;\mu)
\right\}
\nonumber
\\
&=&
\mu_{\rm{N}^\prime}^2
e^{-\mu}
+
\left[
\rule[-0.1em]{0em}{1.2em}
\mu
\right]
-2\mu_{\rm{N}^\prime}
\left[
\rule[-0.1em]{0em}{1.2em}
1-e^{-\mu}
\right]
+\mu_{\rm{N}^\prime}^2
\left\{
\rule[-0.1em]{0em}{1.2em}
e^{-\mu}
\Bigl[
\mbox{Ei}(\mu) - \gamma - \ln(\mu)
\Bigl]
\right\}
\nonumber
\\
&=&
\mu_{\rm{N}^\prime}^2
e^{-\mu}
\Bigl[
1 + \mbox{Ei}(\mu) - \gamma - \ln(\mu)
\Bigl]
-2\mu_{\rm{N}^\prime}
\left[
\rule[-0.1em]{0em}{1.2em}
1-e^{-\mu}
\right]
+
\left[
\rule[-0.1em]{0em}{1.2em}
\mu
\right]
\nonumber
\\
&=&
\mu_{\rm{N}^\prime}^2
e^{-\mu}
\left[
\frac{e^{\mu}-1}{\mu_{\rm{N}^\prime}}
\right]
-2\mu_{\rm{N}^\prime}
\left[
\rule[-0.1em]{0em}{1.2em}
1-e^{-\mu}
\right]
+
\left[
\rule[-0.1em]{0em}{1.2em}
\mu
\right]
\nonumber
\\
&=&
\left\{
\mu_{\rm{N}^\prime}
\right\}
\left[
e^{-\mu}
- 1
\right]
+
\mu
\nonumber
\\
&=&
\left\{
\frac{e^\mu-1}{1 + \mbox{Ei}(\mu) - \gamma - \ln(\mu)}
\right\}
\left[
e^{-\mu}
- 1
\right]
+
\mu
\nonumber
\\
&=&
\frac{\displaystyle
2
-
e^{\mu}
-
e^{-\mu}
}{\displaystyle
{\rm{Ei}}(\mu) - \gamma - \ln(\mu) + 1
}
+
\mu
\label{eq:reduced_x2n}
{}~.
\end{eqnarray}
In the limit of large mean Poisson values,
we see that the numerator of the first term
of Equation
(\ref{eq:reduced_x2n})
is dominated by the $-e^\mu$ term
while the denominator of the first term
is dominated by the Ei$(\mu)$ term
which asymptotically approaches the value of $e^\mu/(\mu-1)$.
We then conclude that
the reduced chi-square of the $\chi^2_{\rm{N}}$ statistic
applied to a Poisson distribution
[Equation (\ref{eq:reduced_x2n})]
approaches
the value of one for large
true Poisson mean values.
Figure \ref{fig-reduced_chi_square}
\firstuse{Fig\ref{fig-reduced_chi_square}}
graphically confirms this finding;
we see that Equation (\ref{eq:reduced_x2n})
reaches a value of $\sim$$1$ only for very
large true Poisson mean values
($\mu \gea 100$).

The reduced chi-square of the new $\chi^2_\gamma$ statistic is, by
definition, the value of the $\chi^2_\gamma$ statistic divided by
the number of degrees of freedom:
\begin{equation}
\frac{
\chi^2_\gamma
}{
\nu
}
\equiv
\frac{1}{N-M}
{\sum_{i=1}^{N}}
\frac{ [n_i + \min(n_i,1) - m_i]^2 }{ n_i + 1 }
\ .
\end{equation}
Now let us assume that our data comes from a Poisson distribution with a
mean value of $\mu$.
In this case, the model $m_i$ will be a constant,
$\mu_\gamma$ [Equation (\ref{eq:weighted_mean_gamma})],
which will,
on average, have a value,
$\mu_{\gamma^\prime}$,
given by
Equation (\ref{eq:weighted_mean_limit_gamma})
in the limit of a large number of observations
(N.B. $\mu_{\gamma^\prime} \equiv \mu$).
Given these assumptions, we find that,
in the limit of a large number of observations,
the reduced chi-square of the new $\chi^2_\gamma$ statistic with
the model $\mu_\gamma$ is
\begin{eqnarray}
\lim_{N \rightarrow \infty}
\left[
\frac{\chi^2_\gamma}{\nu}
\right]
&\equiv&
\lim_{N \rightarrow \infty}
\left[
\frac{1}{N-M}
{\sum_{i=1}^{N}}
\frac{ [n_i + \min(n_i,1) - m_i]^2 }{ n_i + 1 }
\right]
\nonumber
\\
&=&
\lim_{N \rightarrow \infty}
\left[
\frac{1}{N-1}
{\sum_{i=1}^{N}}
\frac{ [n_i + \min(n_i,1) - \mu_\gamma]^2 }{ n_i + 1 }
\right]
\nonumber
\\
&\approx&
\lim_{N \rightarrow \infty}
\left[
\frac{1}{N-1}
{\sum_{k=0}^{\infty}}
\frac{ [k + \min(k,1) - \mu_{\gamma^\prime}]^2 }{ k + 1 }
\left\{
\rule[-0.1em]{0em}{1.2em}
N P(k;\mu)
\right\}
\right]
\nonumber
\\
&=&
{\sum_{k=0}^{\infty}}
\frac{ [k + \min(k,1) - \mu_{\gamma^\prime}]^2 }{ k + 1 }
\left\{
\rule[-0.1em]{0em}{1.2em}
P(k;\mu)
\right\}
\nonumber
\\
&=&
{\sum_{k=0}^{\infty}}
\frac{ [k + \min(k,1) - \mu]^2 }{ k + 1 }
\left\{
\rule[-0.1em]{0em}{1.2em}
P(k;\mu)
\right\}
\nonumber
\\
&=&
\mu^2
e^{-\mu}
+
{\sum_{k=1}^{\infty}}
\frac{ [k + 1 - \mu]^2 }{ k + 1 }
\left\{
\rule[-0.1em]{0em}{1.2em}
P(k;\mu)
\right\}
\nonumber
\\
&=&
\mu^2
e^{-\mu}
+
\left[
\rule[-0.1em]{0em}{1.2em}
{\sum_{k=1}^{\infty}}
k
P(k;\mu)
\right]
+
\left(
1-2\mu
\right)
\left[
\rule[-0.1em]{0em}{1.2em}
{\sum_{k=1}^{\infty}}
P(k;\mu)
\right]
+\mu^2
\left[
\rule[-0.1em]{0em}{1.2em}
{\sum_{k=1}^{\infty}}
\frac{1}{k+1}
P(k;\mu)
\right]
\nonumber
\\
&=&
\mu^2
e^{-\mu}
+
\left[
\rule[-0.1em]{0em}{1.2em}
\mu
\right]
+
\left(
1-2\mu
\right)
\left[
\rule[-0.1em]{0em}{1.2em}
1-e^{-\mu}
\right]
+\mu^2
\left[
\rule[-0.1em]{0em}{1.2em}
\frac{1}{\mu}\left(1-e^{-\mu}\right) - e^{-\mu}
\right]
\nonumber
\\
&=&
1 + e^{-\mu}\left( \mu - 1 \right)
\label{eq:reduced_x2g}
{}~.
\end{eqnarray}
Figure \ref{fig-reduced_chi_square}
shows that
the reduced chi-square of the $\chi^2_{\gamma}$ statistic
applied to a Poisson distribution
[Equation (\ref{eq:reduced_x2g})]
approaches
the value of one for
small true Poisson mean values (i.e.\ $\mu \gea 7$).

Figure \ref{fig-variance_reduced_chi_square}
\firstuse{Fig\ref{fig-variance_reduced_chi_square}}
shows the variance of the reduced chi-square of the
$\chi^2_{\rm{P}}$,
$\chi^2_{\rm{N}}$,
$\chi^2_\gamma$,
and
$\chi^2_\lambda$
statistics
as a function of the true Poisson mean.  This figure was derived
by analyzing the data used in Figure
\ref{fig-weighted_mean_modified_neyman}.

\section{SIMULATED X-RAY POWER-LAW SPECTRA}

I now
demonstrate the new $\chi^2_\gamma$
statistic by using it to study a dataset of simulated X-ray
power-law spectra.  This dataset is based on my
duplication of the simple numerical experiment of
Nousek \& Shue (\cite{nosh1989}).
The number of X-ray photons per energy interval (bin) of a X-ray power-law
spectrum is
\begin{equation}
dN
=
N_0 E^{-\gamma}\,dE
\ .
\end{equation}
Over an energy range,
$E_{\rm{min}} \leq E \leq E_{\rm{max}}$ keV,
the expectation value for the total number of counts can be determined
as follows
\begin{equation}
N
=
N_0
\int_{E_{\rm{min}}}^{E_{\rm{max}}}
E^{-\gamma}\,dE
\ ,
\end{equation}
which implies that
\begin{equation}
N_0
=
\frac{
   N
}{
   {E^{1-\gamma}_{\rm{min}}}
   -
   {E^{1-\gamma}_{\rm{max}}}
}
\label{eq:n0_definition}
\ .
\end{equation}
Following Nousek \& Shue,
I
chose the slope value of
$\gamma \equiv 2.0$
and used the  energy range of 0.095--0.845 keV
which was split into 15 equal bins of 0.050 keV per bin.
I
simulated $10^4$ X-ray spectra for each of the
theoretical $N$ values used by Nousek \& Shue:
25, 50, 75, 100, 150, 250, 500, 750, 1000, 2500, 5000, and $10^4$
photons per spectrum.
Figure \ref{fig-spectra}
\firstuse{Fig\ref{fig-spectra}}
shows four of the simulated X-ray power-law spectra.

\subsection{Powell's Method:
Solving for $\gamma$ and $N$ using
$\chi^2_{\rm{N}}$,
$\chi^2_{\rm{P}}$,
$\chi^2_\gamma$
}

I
determined the best-fit model parameters
$\gamma_{\rm{calc}}$ and $N_{\rm{calc}}$
for each simulated spectrum with
Powell's function minimization method\footnote{
The primary reference for
Powell's minimization method is
Powell (\cite{po1964}).
More accessible descriptions may be found
in the numerical-methods literature
(e.g.,
Acton \cite{ac1970},
Gill, Murray \& Wright \cite{giet1981},
and
Press \et \cite{pret1986})
}
using the modified Neyman's $\chi^2$ statistic ($\chi^2_{\rm{N}}$),
Pearson's $\chi^2$ statistic ($\chi^2_{\rm{P}}$),
and the new $\chi^2_\gamma$ statistic.
I
used the following crude initial guesses:
$\gamma=0.0$ and $N=1.3\sum_i^{15} n_i$,
where $n_i$ is the observed
number of photons in the $i$th channel (bin).
I
computed the robust mean (average) and robust standard deviation\footnote{
The robust mean given in all the tables
is the mean of all values within two
average deviations of the standard mean value.
The robust standard deviation given in all the tables
is $1.55$$\sigma$
where $\sigma$ is the standard deviation of all values within
two average deviations of the standard mean values.
}
of the ratios $\gamma_{\rm{calc}}/\gamma$ and $N_{\rm{calc}}/N$ for the
$10^4$ simulated spectra of each dataset.
The results of Powell's method with two free parameters
($\gamma,N$) using the
$\chi^2_{\rm{N}}$,
$\chi^2_{\rm{P}}$,
$\chi^2_\gamma$
statistics are presented in
Table \ref{tbl-powell_2fp}
\firstuse{Tab\ref{tbl-powell_2fp}}
and
Figure \ref{fig-powell_2fp}
\firstuse{Fig\ref{fig-powell_2fp}}
{}.
The first column,
$N$,
of Table \ref{tbl-powell_2fp}
corresponds to the total theoretical number of counts in the spectrum.
The columns
``$\gamma_{\rm{calc}}/\gamma$''
and
``$N_{\rm{calc}}/N$''
are the robust mean values of the ratios of the best-fit parameters divided by
the original value that was used to create the datasets.
The parenthetical numbers are the robust standard deviations
which can be used to determine the significance of the deviation from
the perfect ratio value of one.
For example, the first value of the 2nd column of Table \ref{tbl-powell_2fp}
is 1.002(11) which represents the value of $1.002$$\pm$$0.011$.
The deviation of this value from one (i.e. $0.002$) is
statistically significant since the error of the mean is only
$\sim$$0.011/\sqrt{10^4}$ or $\sim$$0.00011$.

Figure \ref{fig-powell_2fp}  indicates that
the new $\chi^2_\gamma$ statistic gives the best results.
Using a 5\% criteria for both fitted parameters $(\gamma,N)$,
we see that the $\chi^2_\gamma$ statistic gives
good results for spectra with $\gea$$50$ photons.
By comparison,
Pearson's $\chi^2$ statistic requires $\gea$$250$ photons
and the modified Neyman's $\chi^2$ statistic requires $\gea$$750$ photons
in order to get the same quality of results.
Baker \& Cousins (\cite{baco1984})
noted that, in many cases, $\chi^2$ fits using the
the modified Neyman's $\chi^2$ statistic
will underestimate the total number of counts
while $\chi^2$ fits using
Pearson's $\chi^2$ statistic will overestimate the total number of counts;
both systematic errors are clearly seen in
the bottom panel of Figure \ref{fig-powell_2fp}.
I stated in the previous section that the usage of the modified Neyman's
$\chi^2$ statistic with Poisson-distributed data
will typically underestimate the total counts
by one count per bin.
My results for the $\chi^2_{\rm{N}}$ statistic
clearly exhibit this systematic error:
the results of the
ratio $N_{\rm{calc}}/N$ for spectra with $N \gea 250$ photons
(squares in the bottom panel of Fig.\ \ref{fig-powell_2fp})
are well modeled by the function $(N-15)/N$
where 15 is the number of bins (channels) in our spectra
[see the dashed curve in the bottom panel of Fig.\ \ref{fig-powell_2fp}].

A comparison of my
analysis of $\gamma_{\rm{calc}}/\gamma$ using
the modified Neyman's $\chi^2$ statistic
(2nd column of Table \ref{tbl-powell_2fp})
with the analysis of Nousek \& Shue for Powell's method
(3rd column of their Table 3)
shows nearly identical results.
In my
version of this numerical experiment,
I
used the two parameters $N$ and $\gamma$
while Nousek \& Shue used $N_0$ and $\gamma$.
A comparison of my
analysis of $N_{\rm{calc}}/N$
(3rd column of Table \ref{tbl-powell_2fp})
with their Powell's method analysis of $N_{\rm{calc}}/N_0$
(2nd column of their Table 3)
shows that my
analysis with
$N_{\rm{calc}}/N$
has produced better estimates.
This should not be surprising because the parameter $N_0$ is
not an independent parameter -- $N_0$ depends on both the
slope of the spectrum and the theoretical number
of photons in the spectrum.
As a general rule, one gets better results by solving for
independent parameters instead of dependent parameters.

\subsection{Levenberg-Marquardt Method:
Solving for $\gamma$ and $N$ using
$\chi^2_{\rm{N}}$,
$\chi^2_{\rm{P}}$,
$\chi^2_\gamma$
}

I
determined the best-fit model parameters
$\gamma_{\rm{calc}}$ and $N_{\rm{calc}}$
for each simulated spectrum with
Levenberg-Marquardt method\footnote{
The primary references for
Levenberg-Marquardt method are
Levenberg (\cite{le1944})
and
Marquardt (\cite{ma1963}).
More accessible descriptions may be found
in the numerical-methods literature
(e.g.,
Bevington \cite{be1969},
Gill, Murray \& Wright \cite{giet1981},
and
Press \et \cite{pret1986})
}
using the modified Neyman's $\chi^2$ statistic ($\chi^2_{\rm{N}}$),
Pearson's $\chi^2$ statistic ($\chi^2_{\rm{P}}$),
and the new $\chi^2_\gamma$ statistic.
I
used the previous crude initial guesses:
$\gamma=0.0$ and $N=1.3\sum_i^{15} n_i$.
I
computed the robust mean
and robust standard deviation
of the ratios $\gamma_{\rm{calc}}/\gamma$ and $N_{\rm{calc}}/N$ for the
$10^4$ simulated spectra of each dataset.
The results of Levenberg-Marquardt method with two free parameters
($\gamma,N$) using the
$\chi^2_{\rm{N}}$,
$\chi^2_{\rm{P}}$,
$\chi^2_\gamma$
statistics are presented in
Table \ref{tbl-marquardt_2fp}
\firstuse{Tab\ref{tbl-marquardt_2fp}}
and
Figure \ref{fig-marquardt_2fp}
\firstuse{Fig\ref{fig-marquardt_2fp}}
{}.

Figure \ref{fig-marquardt_2fp}  indicates that
the new $\chi^2_\gamma$ statistic gives the best results.
Using a 5\% criteria for both fitted parameters $(\gamma,N)$,
we see that the $\chi^2_\gamma$ statistic gives
good results for all the spectra ($N\gea25$ photons).
By comparison,
Pearson's $\chi^2$ statistic requires $\gea$$100$ photons
and the modified Neyman's $\chi^2$ statistic requires $\gea$$500$ photons
in order to get the same quality of results.

The results for the $\chi^2_\gamma$ and $\chi^2_{\rm{N}}$ statistics
are nearly identical with either Powell's method
(Table \ref{tbl-powell_2fp})
or the Levenberg-Marquardt method
(Table \ref{tbl-marquardt_2fp}).
This finding refutes the determination by
Nousek \& Shue (\cite{nosh1989}) that
Powell's method gives more accurate results than
the Levenberg-Marquardt method.

The results for Pearson's $\chi^2$ improved significantly by
using the Levenberg-Marquardt method instead of Powell's method.
An inspection of the
individual fits showed that the Levenberg-Marquardt
method with the $\chi^2_{\rm{P}}$ statistic produced a best-fit
value for $N$ that was within a one-tenth of one percent of
the total number of photons in the spectrum.
Needless to say,
with such an improvement in the determination of $N$,
a much better estimate for the slope $\gamma$ could be determined.

This peculiar result tells us something important about this
particular minimization problem: an excellent estimate of the
total number of photons in the best-fit spectrum is the total
number of photons in the actual spectrum.
Thus by setting $N$ to be a constant,
$N\equiv\sum^{15}_i n_i$, we can eliminate one parameter
and solve for $\gamma$ alone.

\subsection{Powell's Method:
Solving for $\gamma$ using
$\chi^2_{\rm{N}}$,
$\chi^2_{\rm{P}}$,
$\chi^2_\gamma$
}

I
determined the best-fit model parameter
$\gamma_{\rm{calc}}$
for each simulated spectrum with
Powell's function minimization method
using the modified Neyman's $\chi^2$ statistic ($\chi^2_{\rm{N}}$),
Pearson's $\chi^2$ statistic ($\chi^2_{\rm{P}}$),
and the new $\chi^2_\gamma$ statistic.
I
set $N\equiv\sum_i^{15} n_i$ and
used the crude initial guess of
$\gamma=0.0$.
I
computed the robust mean
and robust standard deviation
of the ratios $\gamma_{\rm{calc}}/\gamma$ for the
$10^4$ simulated spectra of each dataset.
The results of Powell's method with two free parameters
($\gamma,N$) using the
$\chi^2_{\rm{N}}$,
$\chi^2_{\rm{P}}$,
$\chi^2_\gamma$
statistics are presented in
Table \ref{tbl-powell_1fp}
\firstuse{Tab\ref{tbl-powell_1fp}}
and
Figure \ref{fig-powell_1fp}
\firstuse{Fig\ref{fig-powell_1fp}}
{}.

Figure \ref{fig-powell_1fp}  indicates that
the new $\chi^2_\gamma$ statistic gives the best results.
Using a 5\% criteria,
we see that the $\chi^2_\gamma$ statistic gives
good results for all the spectra ($N\gea25$ photons).
By comparison,
Pearson's $\chi^2$ statistic requires $\gea$$250$ photons
and the modified Neyman's $\chi^2$ statistic requires $\gea$$750$ photons
in order to get the same quality of results.

Fitting only for the slope $\gamma$ has improved the results for
the new $\chi^2_\gamma$ statistic and
the modified Neyman's $\chi^2$ statistic.
The results for Pearson's $\chi^2$ show no improvement over the
two free parameter result.

\subsection{Levenberg-Marquardt Method:
Solving for $\gamma$ using
$\chi^2_{\rm{N}}$,
$\chi^2_{\rm{P}}$,
$\chi^2_\gamma$
}

I
determined the best-fit model parameter
$\gamma_{\rm{calc}}$
for each simulated spectrum with
the Levenberg-Marquardt minimization method
using the modified Neyman's $\chi^2$ statistic ($\chi^2_{\rm{N}}$),
Pearson's $\chi^2$ statistic ($\chi^2_{\rm{P}}$),
and the new $\chi^2_\gamma$ statistic.
I
set $N\equiv\sum_i^{15} n_i$ and
used the crude initial guess of
$\gamma=0.0$.
I
computed the robust mean
and robust standard deviation
of the ratios $\gamma_{\rm{calc}}/\gamma$ for the
$10^4$ simulated spectra of each dataset.
The results of the Levenberg-Marquardt method with one free parameter
($\gamma$) using the
$\chi^2_{\rm{N}}$,
$\chi^2_{\rm{P}}$,
$\chi^2_\gamma$
statistics are presented in
Table \ref{tbl-marquardt_1fp}
\firstuse{Tab\ref{tbl-marquardt_1fp}}
and
Figure \ref{fig-marquardt_1fp}
\firstuse{Fig\ref{fig-marquardt_1fp}}
{}.

Figure \ref{fig-marquardt_1fp}  indicates that
Pearson's $\chi^2$ statistic gives the best results.
Using a 5\% criteria,
we see that both
the new $\chi^2_\gamma$ statistic
and
the $\chi^2_{\rm{P}}$ statistic
give good results for all the spectra ($N\gea25$ photons).
By comparison,
the modified Neyman's $\chi^2$ statistic still requires $\gea$$750$ photons
in order to get the same quality of results.
Once again, we note that
the results for the $\chi^2_\gamma$ and $\chi^2_{\rm{N}}$ statistics
are nearly identical with either Powell's method
(Table \ref{tbl-powell_1fp})
or the Levenberg-Marquardt method
(Table \ref{tbl-marquardt_1fp}).

\subsection{Error Estimates}

One expects the quality of the slope determination to degrade
as the total number of photons in the X-ray spectra decline.
Figure \ref{fig-marquardt_powell_1fp}
\firstuse{Fig\ref{fig-marquardt_powell_1fp}}
shows the distribution of the best-fit
values for the slope $\gamma$ for the faintest
spectra with a
theoretical
total of
100, 50, and 25 photons.
As expected, the range of best-fit slope values measured for
spectra with only $N\equiv25$ photons is considerably larger than
the range of values for spectra with $N\equiv100$ photons.

The Levenberg-Marquardt method not only provides best-fit values
for parameters but it also provides an error estimate
(approximately $1\,\sigma$ errors)
of those fitted parameters.
How believable are these error estimates?
Figure \ref{fig-error_estimates}
\firstuse{Fig\ref{fig-error_estimates}}
shows an analysis of the errors
estimated by the Levenberg-Marquardt method when
the new $\chi^2_\gamma$ statistic was used
to analyze spectra with theoretical totals
of 100, 50, and 25 photons.

The top panel
of Figure \ref{fig-error_estimates}
shows the error analysis of
spectra with $N\equiv100$ photons.
The median slope value is 1.989
and the median error estimate is 0.194\,.
A total of 15.87\% of the spectra have
estimates of $\gamma \leq 1.789$ and
15.87\% of the spectra have
estimates of $\gamma \geq 2.211$.
For a normal distribution, one expects 68.26\% of the deviates to be
found within one standard deviation of the mean.
Assuming that the
distribution of best-fit $\gamma$ values approximates a normal
distribution,
then half of the difference between
the 84.13 and 15.87 percentile values of $\gamma$ can be used
as an estimate for the slope error:
$\sigma_\gamma
\approx
(\gamma_{84.13\%} - \gamma_{15.87\%})/2
=
(2.211-1.789)/2
=
0.211$\,.
This value is 8.8\% larger than the median Levenberg-Marquardt error
estimate; a fractional error of 10.6\% instead of the predicted
9.8\%\,.

The middle panel
of Figure \ref{fig-error_estimates}
shows the error analysis of
spectra with $N\equiv50$ photons.
The median slope value is 2.009
and the median error estimate is 0.301\,.
A total of 15.87\% of the spectra have
estimates of $\gamma \leq 1.732$ and
15.87\% of the spectra have
estimates of $\gamma \geq 2.334$.
This gives an estimated slope error of
$\sigma_\gamma
\approx
(\gamma_{84.13\%} - \gamma_{15.87\%})/2
=
0.301$\,.
This value is exactly equal to the median Levenberg-Marquardt error
estimate.

The bottom panel
of Figure \ref{fig-error_estimates}
shows the error analysis of
spectra with $N\equiv25$ photons.
The median slope value is 2.071
and the median error estimate is 0.484\,.
A total of 15.87\% of the spectra have
estimates of $\gamma \leq 1.692$ and
15.87\% of the spectra have
estimates of $\gamma \geq 2.570$.
This gives an estimated slope error of
$\sigma_\gamma
\approx
(\gamma_{84.13\%} - \gamma_{15.87\%})/2
=
0.439$\,.
This value is 9.3\% less than the median Levenberg-Marquardt error
estimate; a fractional error of 21.2\% instead of the predicted
23.4\%\,.

{\em{The errors estimated by the Levenberg-Marquardt method are seen
to be reasonable.}}
Figure \ref{fig-spectra_fits}
\firstuse{Fig\ref{fig-spectra_fits}}
shows the simulated X-ray spectra of Fig.\ \ref{fig-spectra}
now plotted with $\chi^2_\gamma$ fits produced by the Levenberg-Marquardt
method with one free parmater.  The Levenberg-Marquardt method has done
a good job even with the two faintest spectra which have actual
totals of only 28 and 101 photons.

\subsection{The $\chi^2_\lambda$ and Cash's $C$ statistics}

For the sake of completeness,
I
determined the best-fit model parameter
$\gamma_{\rm{calc}}$
for each simulated spectrum with
Powell's function minimization method
using the maximum likelihood ratio statistic for Poisson distributions,
$\chi^2_\lambda$ [Equation (\ref{eq:x2l})\,],
and Cash's $C$ statistic,
\begin{equation}
C
\equiv
2{\sum_{i=1}^{N}}
\left[
m_i - n_i\ln\left({m_i}\right)
\right]
\end{equation}
[Equation (6) of Cash \cite{ca1979}].
I
set $N\equiv\sum_i^{15} n_i$ and
used the crude initial guess of
$\gamma=0.0$.
I
computed the robust mean
and robust standard deviation
of the ratios $\gamma_{\rm{calc}}/\gamma$ for the
$10^4$ simulated spectra of each dataset.
The results of Powell's method with one free parameter
($\gamma$) using the
$\chi^2_\lambda$ statistic
and Cash's $C$ statistic
are presented in
Table \ref{tbl-powell_1fp_x2l}
\firstuse{Tab\ref{tbl-powell_1fp_x2l}}
and
Figure \ref{fig-x2l}
\firstuse{Fig\ref{fig-x2l}}
{}.

Table \ref{tbl-powell_1fp_x2l} and the right panel of
Figure \ref{fig-x2l}
shows that Cash's $C$ statistic and
the maximum likelihood ratio statistic for Poisson distributions,
$\chi^2_\lambda$,
give identical results.
This is not surprising because Cash's $C$ statistic
is a variant of the more well-known $\chi^2_\lambda$ statistic
which has been discussed in the literature for over 70 years
(e.g., Neyman \& Pearson \cite{nepe1928}).

I
also determined the best-fit model parameter
$\gamma_{\rm{calc}}$
for each simulated spectrum with
the Levenberg-Marquardt minimization method
using the maximum likelihood ratio statistic for Poisson distributions,
$\chi^2_\lambda$.
I
set $N\equiv\sum_i^{15} n_i$ and
used the crude initial guess of
$\gamma=0.0$.
I
computed the robust average and robust standard deviation
of the ratios $\gamma_{\rm{calc}}/\gamma$ for the
$10^4$ simulated spectra of each dataset.
The results of the Levenberg-Marquardt  method with one free parameter
($\gamma$) using the
$\chi^2_\lambda$ statistic
is presented in
Table \ref{tbl-marquardt_1fp_x2l}
\firstuse{Tab\ref{tbl-marquardt_1fp_x2l}}
and
Figure \ref{fig-x2l} .
The maximum likelihood ratio statistic for Poisson distributions,
$\chi^2_\lambda$,
produces nearly identical results with either Powell's method or
the Levenberg-Marquardt minimization method.

Of the two statistics, $\chi^2_\lambda$ and the new $\chi^2_\gamma$,
which is better?
Although Tables
\ref{tbl-marquardt_1fp_x2l}
and
\ref{tbl-marquardt_1fp}
indicate that the $\chi^2_\lambda$ is slightly better,
we see that the actual differences between the distributions presented
in Figure
\ref{fig-x2l}
are really quite negligible when compared with the overall
uncertainty caused by simple sampling errors (counting statistics)
of the simulated X-ray spectra.

\section{SUMMARY}

I
have demonstrated that the application of
the standard weighted mean formula,
$
\left[\sum_i {n_i \sigma^{-2}_i}\right]
/
\left[\sum_i {\sigma^{-2}_i}\right]
$,
to determine the weighted mean
of data, $n_i$, drawn from a Poisson distribution, will,
on average,
underestimate the true mean by $\sim$$1$ for all true mean
values larger than $\sim$$3$
when the common assumption is made
that the error of the $i$th observation is
$\sigma_i = \max(\sqrt{n_i},1)$.
This small, but statistically significant offset,
explains the long-known observation that chi-square minimization techniques
which use the modified Neyman's $\chi^2$ statistic,
$\chi^2_{\rm{N}} \equiv \sum_i (n_i-y_i)^2/\max(n_i,1)$,
to compare Poisson-distributed data with model values, $y_i$, will
typically predict a total number of counts that
underestimates the true total
by about $1$ count per bin.
Based on my
finding that the weighted mean of data
drawn from a Poisson distribution can be
determined using the formula
$
\left[
\sum_i \left[n_i+\min\left(n_i,1\right)\right]\left(n_i+1\right)^{-1}
\right]
/
\left[
\sum_i \left(n_i+1\right)^{-1}
\right]
$, I
proposed that a new $\chi^2$ statistic,
$\chi^2_\gamma
\equiv
\sum_i
\left[ n_i + \min\left( n_i, 1\right) - y_i\right]^2
/
\left[ n_i + 1 \right]$,
should always be used to analyze Poisson-distributed data
in preference to the modified Neyman's $\chi^2$ statistic.

I
demonstrated the power and usefulness of $\chi^2_\gamma$ minimization
by using two statistical fitting techniques
(Powell's method and the Levenberg-Marquardt method)
and five $\chi^2$ statistics
($\chi^2_{\rm{N}}$,
$\chi^2_{\rm{P}}$,
$\chi^2_\gamma$,
$\chi^2_\lambda$,
and Cash's $C$\,)
to analyze simulated X-ray power-law 15-channel spectra
with large and small counts per bin.
I
showed that $\chi^2_\gamma$ minimization with
the Levenberg-Marquardt or Powell's method can produce
excellent results (mean slope errors $\lea$$3$\%)
with spectra having as few as 25 total counts.

This analysis shows that there is nothing inherently wrong with either
the Levenberg-Marquardt method or Powell's method in the low-count regime
--- provided that one uses an appropriate $\chi^2$ statistic for the
type of data being analyzed.  Given Poisson-distributed data,
one should always use the new $\chi^2_\gamma$ statistic in preference
to the modified Neyman's $\chi^2$ statistic because
that statistic produces small, but statistically significant,
systematic errors with Poisson-distributed data.

While the new $\chi^2_\gamma$ statistic is not perfect,
neither is the more well-known $\chi^2_\lambda$ statistic
(e.g., see Figures
\ref{fig-reduced_chi_square}
and
\ref{fig-variance_reduced_chi_square}).
Both statistics have problems in the very-low-count regime.
The new $\chi^2_\gamma$ statistic complements but does not replace the
older $\chi^2_\lambda$ statistic.
Which statistic is ``best'' will generally depend on the particular problem
being analyzed.
An important difference between these two statistics
is that the $\chi^2_\lambda$ statistic assumes that all data is perfect.
With data from perfect counting experiments,
the $\chi^2_\lambda$ statistic may give slightly better results than the new
$\chi^2_\gamma$ statistic.  However, data is typically obtained under
less-than-perfect circumstances with multiple imperfect detectors.
The $\chi^2_\gamma$ statistic, by definition, is a {\em{weighted}} $\chi^2$
statistic which makes it easy to assign a {\em{lower}} weight to data
from poor detectors.  Thus in the analysis of real data obtained
with noisy and imperfect detectors, the $\chi^2_\gamma$ statistic may well
outperform the classic $\chi^2_\lambda$ statistic because low-quality data
can be given a lower weight instead of being completely rejected.

Finally, I
note in passing that two simple transformations
may make it possible to retrofit
many existing computer implementations (i.e.\ executable binaries)
of $\chi^2_{\rm{N}}$ minimization algorithms
to do $\chi^2_\gamma$ minimization through the simple expedient
of {\em{changing the input data}}
from
$\left[n_i\right]$
to
$\left[n_i + \min\left(n_i,1\right)\right]$,
and error estimates,
$\sigma_i$,
from
$\left[\max\left(\sqrt{n_i},1\right)\right]$
to
$\left[\sqrt{n_i+1}\right]$.

\acknowledgments

I would like to thank my former colleagues at Mount Stromlo and Siding Spring
Observatories and current colleages at the
National Optical Astronomy Observatories
with whom I have had many useful discussions about various
topics related to this research.
Keith Arnaud, Martin Elvis, and Andy Ptak provided valuable practical
information about the implementation details of the $\chi^2_\lambda$ statistic
within the {\sc{xspec}} X-ray spectral-fitting program.
I would also like to thank the anonymous referee whose thoughtful comments and
suggestions improved the final article.
I am grateful to the anonymous referee
of the previous version of
this paper (based on the $\chi^2_\beta$ statistic)
for providing a copy of the Baker \& Cousins (\cite{baco1984}) article.
Special thanks are due to
Christopher Burke who checked all the proofs in an early draft of the
manuscript
and to Ian Dell'Antonio who carefully read the second-to-last revision
of the manuscript.
Mary Guerrieri has my heartfelt thanks for cheerfully helping me acquire
copies of the many articles in statistical journals which are not
available in the NOAO library collection.
KJM
was supported by a grant from
the National Aeronautics and Space Administration (NASA),
Order No.\ S-67046-F, which was awarded by
the Long-Term Space Astrophysics Program (NRA 95-OSS-16).
Publication costs were partially paid by NASA through Grant No.~STSCI GO-5386
from the Space Telescope Science Institute, which is operated by the
by the Association of Universities for Research in Astronomy, Inc., under
NASA Contract No.~NAS5-26555.
This research has made use of
NASA's Astrophysics Data System Abstract Service
which is operated by the Jet Propulsion Laboratory at
the California Institute of Technology,
under contract with NASA.

\def\fig01cap{
\label{fig-weighted_mean_modified_neyman}
\noteforeditor{Print this figure TWO (2) COLUMNS wide.\newline}
Analysis of four weighted-mean formulae applied to Poisson-distributed data.
Each open square represents the weighted mean of $4$$\times$$10^6$
Poisson deviates at each given true mean value:
$ 0.001 < \mu < 1000$.
\newline
{\bf{(a)}}
The difference between the weighted mean computed using
Equation (\protect\ref{eq:weighted_mean_modified_neyman}),
$\mu_{\rm{N}}$,
and the true mean,
$\mu\,$.
The solid curve is the difference between
Equation (\protect\ref{eq:wmean_limit_neyman}) and the true mean:
$\left\{[e^{\mu}-1][1+{\rm{Ei}}(\mu)-\gamma-\ln(\mu)]^{-1}\right\}-\mu.$
\newline
{\bf{(b)}}
The difference between the weighted mean computed using
Equation (\protect\ref{eq:weighted_mean_alpha}),
$\mu_\alpha$,
and the true mean,
$\mu\,$.
The solid curve is the difference between
Equation (\protect\ref{eq:weighted_mean_limit_alpha})
and the true mean:
$\left\{\mu[1-e^{-\mu}]^{-1}-1\right\}-\mu$.
\newline
{\bf{(c)}}
The difference between the weighted mean computed using
Equation (\protect\ref{eq:weighted_mean_beta}),
$\mu_\beta$,
and the true mean,
$\mu\,$.
The solid curve is the difference between
Equation (\protect\ref{eq:weighted_mean_limit_beta})
and the true mean:
$\left\{\mu[1-e^{-\mu}]^{-1}\right\}-\mu$.
\newline
{\bf{(d)}}
The difference between the weighted mean computed using
Equation (\protect\ref{eq:weighted_mean_gamma}),
$\mu_\gamma$,
and the true mean,
$\mu\,$.
The solid curve is the difference between
Equation (\protect\ref{eq:weighted_mean_limit_gamma})
and the true mean.
The difference is zero because
$\mu_\gamma$
is the weighted-mean formula for Poisson-distributed data.
}

\ifundefined{showfigs}{
  \newpage
  \centerline{{\Large\bf{Figure Captions}}}
  \smallskip
  \figcaption{\fig01cap}
}\else{
  \clearpage
  \newpage
  \begin{figure}
  \figurenum{1}
  \vspace*{-35truemm}
  \hspace*{-0truemm}
  \epsfxsize=6.5truein
  \epsfbox{mighell.fig01.eps}
  \vspace*{-40truemm}
  \caption[]{\baselineskip 1.15em \fig01cap}
  \end{figure}
}
\fi

\def\fig02cap{
\label{fig-reduced_chi_square}
\noteforeditor{Print this figure ONE (1) COLUMN wide.\newline}
Reduced chi-square
$(\chi^2/\nu)$
as a function of true Poisson mean,
$\mu$,
for 4 $\chi^2$ statistics:
Pearson's $\chi^2$
$[
\chi^2_{\rm{P}}
\equiv
{\sum_{i=1}^{N}}
(n_i - m_i)^2/m_i
]$,
the modified Neyman's $\chi^2$
$[
\chi^2_{\rm{N}}
\equiv
{\sum_{i=1}^{N}}
(n_i - m_i)^2/\max(n_i,1)
]$,
the new $\chi^2_\gamma$ statistic
$[
\chi^2_\gamma
\equiv
{\sum_{i=1}^{N}}
(n_i + \min(n_i,1)
 - m_i)^2/(n_i+1)
]$,
and the maximum likelihood ratio statistic for Poisson distributions
$\left[
\chi^2_\lambda
\equiv
2{\sum_{i=1}^{N}}
\left(
m_i - n_i + n_i\ln\left({n_i}/{m_i}\right)
\right)
\right]
$.
The Poisson distributions of
Figure \protect\ref{fig-weighted_mean_modified_neyman}
were analyzed to produce this plot.
The formula for the curve
connecting the values for modified Neyman's $\chi^2$ statistic
($\chi^2_{\rm{N}}$)
is given in Equation (\protect\ref{eq:reduced_x2n}).
The formula for the curve
connecting the values for new $\chi^2_\gamma$ statistic
is given in Equation (\protect\ref{eq:reduced_x2g}).
The dotted line shows the ideal value of one.
}

\ifundefined{showfigs}{
  \figcaption{\fig02cap}
}\else{
  \clearpage
  \newpage
  \begin{figure}
  \figurenum{2}
  \vspace*{-35truemm}
  \hspace*{-0truemm}

  \epsfxsize=6.5truein
  \epsfbox{mighell.fig02.eps}
  \vspace*{-40truemm}
  \caption[]{\baselineskip 1.15em \fig02cap}
  \end{figure}
}
\fi

\def\fig03cap{
\label{fig-variance_reduced_chi_square}
\noteforeditor{Print this figure ONE (1) COLUMN wide.\newline}
The variance of the reduced chi-square
$(\sigma^2_{\chi^2/\nu})$
as a function of true Poisson mean,
$\mu$,
for 5 $\chi^2$ statistics:
the standard $\chi^2$,
Pearson's $\chi^2$
$(\chi^2_{\rm{P}})$,
the modified Neyman's $\chi^2$
$(\chi^2_{\rm{N}})$,
the new $\chi^2_\gamma$ statistic,
and the maximum likelihood ratio statistic for Poisson distributions
$(\chi^2_\lambda)$.
The Poisson distributions of
Figure \protect\ref{fig-weighted_mean_modified_neyman}
were analyzed to produce this plot.
The formula for the variance of the reduced Pearson's $\chi^2$ statistic
is $2+\mu^{-1}$.
The dotted line shows the ideal value of two.
}

\ifundefined{showfigs}{
  \figcaption{\fig03cap}
}\else{
  \clearpage
  \newpage
  \begin{figure}
  \figurenum{3}
  \vspace*{-35truemm}
  \hspace*{-0truemm}

  \epsfxsize=6.5truein
  \epsfbox{mighell.fig03.eps}
  \vspace*{-40truemm}
  \caption[]{\baselineskip 1.15em \fig03cap}
  \end{figure}
}
\fi

\def\fig04cap{
\label{fig-spectra}
\noteforeditor{Print this figure ONE (1) COLUMN wide.\newline}
The dashed lines show 4 ideal X-ray power-law spectra with a total of
25,
100,
1000,
and
10000
photons.
Four simulated X-ray spectra with totals of
28,
101,
1015,
and
9938
photons are shown with 1\,$\sigma$ error bars
estimated with Equations (9) and (14) of Gehrels (\protect\cite{ge1986}).
(N.B. Some errorbars overlap and the bottom two spectra have identical
data values at the 0.47 and 0.67 keV bins.)
}

\ifundefined{showfigs}{
  \figcaption{\fig04cap}
}\else{
  \clearpage
  \newpage
  \begin{figure}
  \figurenum{4}
  \vspace*{-35truemm}
  \hspace*{-0truemm}

  \epsfxsize=6.5truein
  \epsfbox{mighell.fig04.eps}
  \vspace*{-40truemm}
  \caption[]{\baselineskip 1.15em \fig04cap}
  \end{figure}
}
\fi

\def\fig05cap{
\label{fig-powell_2fp}
\noteforeditor{Print this figure ONE (1) COLUMN wide.\newline}
Results of Powell's method
with two free parameters ($\gamma,N$)
for three statistics:
$\chi^2_\gamma$ (circles),
$\chi^2_{\rm{N}}$ (squares),
and
$\chi^2_{\rm{P}}$ (triangles).
This figure uses the data given in Table \protect\ref{tbl-powell_2fp}.
The dotted lines show the ideal ratio value of one.
The dashed curve in the bottom panel shows the function
$(N-15)/N$ which is a good model for the
$N_{\rm{calc}}/N$ results of the
$\chi^2_{\rm{N}}$ statistic for all spectra with $N\gea250$ photons.
}

\ifundefined{showfigs}{
  \figcaption{\fig05cap}
}\else{
  \clearpage
  \newpage
  \begin{figure}
  \figurenum{5}
  \vspace*{-35truemm}
  \hspace*{-0truemm}
  \epsfxsize=6.5truein
  \vspace*{-40truemm}
  \epsfbox{mighell.fig05.eps}
  \caption[]{\baselineskip 1.15em \fig05cap}
  \end{figure}
}
\fi

\def\fig06cap{
\label{fig-marquardt_2fp}
\noteforeditor{Print this figure ONE (1) COLUMN wide.\newline}
Results of the Levenberg-Marquardt method
with two free parameters ($\gamma,N$)
for three statistics:
$\chi^2_\gamma$ (circles),
$\chi^2_{\rm{N}}$ (squares),
and
$\chi^2_{\rm{P}}$ (triangles).
This figure uses the data given in Table \protect\ref{tbl-marquardt_2fp}.
The dotted lines show the ideal ratio value of one.
The dashed curve in the bottom panel shows the function
$(N-15)/N$ which is a good model for the
$N_{\rm{calc}}/N$ results of the
$\chi^2_{\rm{N}}$ statistic for all spectra with $N\gea250$ photons.
}

\ifundefined{showfigs}{
  \figcaption{\fig06cap}
}\else{
  \clearpage
  \newpage
  \begin{figure}
  \figurenum{6}
  \vspace*{-35truemm}
  \hspace*{-0truemm}
  \epsfxsize=6.5truein
  \vspace*{-40truemm}
  \epsfbox{mighell.fig06.eps}
  \caption[]{\baselineskip 1.15em \fig06cap}
  \end{figure}
}
\fi

\def\fig07cap{
\label{fig-powell_1fp}
\noteforeditor{Print this figure ONE (1) COLUMN wide.\newline}
Results of Powell's method
with one free parameter ($\gamma$)
for three statistics:
$\chi^2_\gamma$ (circles),
$\chi^2_{\rm{N}}$ (squares),
and
$\chi^2_{\rm{P}}$ (triangles).
This figure uses the data given in Table \protect\ref{tbl-powell_1fp}.
The dotted lines show the ideal ratio value of one.
}

\ifundefined{showfigs}{
  \figcaption{\fig07cap}
}\else{
  \clearpage
  \newpage
  \begin{figure}
  \figurenum{7}
  \hspace*{-0truemm}
  \epsfxsize=6.5truein
  \vspace*{-100truemm}
  \epsfbox{mighell.fig07.eps}
  \caption[]{\baselineskip 1.15em \fig07cap}
  \end{figure}
}
\fi

\def\fig08cap{
\label{fig-marquardt_1fp}
\noteforeditor{Print this figure ONE (1) COLUMN wide.\newline}
Results of the Levenberg-Marquardt method
with one free parameter ($\gamma$)
for three statistics:
$\chi^2_\gamma$ (circles),
$\chi^2_{\rm{N}}$ (squares),
and
$\chi^2_{\rm{P}}$ (triangles).
This figure uses the data given in Table \protect\ref{tbl-marquardt_1fp}.
The dotted lines show the ideal ratio value of one.
}

\ifundefined{showfigs}{
  \figcaption{\fig08cap}
}\else{
  \clearpage
  \newpage
  \begin{figure}
  \figurenum{8}
  \vspace*{-0truemm}
  \hspace*{-0truemm}
  \epsfxsize=6.5truein
  \vspace*{-100truemm}
  \epsfbox{mighell.fig08.eps}
  \caption[]{\baselineskip 1.15em \fig08cap}
  \end{figure}
}
\fi

\def\fig09cap{
\label{fig-marquardt_powell_1fp}
\noteforeditor{Print this figure TWO (2) COLUMNS wide.\newline}
A comparison of the results of
the analysis of the simulated X-ray spectra with theoretical totals
of 100, 50, and 25 photons using
the Levenberg-Marquardt method
with 1 free parameter (left)
and Powell's method
with 1 free parameter (right).
Note that the histograms for the $\chi^2_\gamma$ and $\chi^2_{\rm{N}}$
are nearly
identical
for both methods.
The statistical analysis of this data is presented in
Tables
\protect\ref{tbl-marquardt_1fp}
and
\protect\ref{tbl-powell_1fp}.
}

\ifundefined{showfigs}{
  \figcaption{\fig09cap}
}\else{
  \clearpage
  \newpage
  \begin{figure}
  \figurenum{9}
  \vspace*{-20truemm}
  \hspace*{-10truemm}
  \epsfxsize=7.0truein
  \epsfbox{mighell.fig09.eps}
  \vspace*{-40truemm}
  \caption[]{\baselineskip 1.15em \fig09cap}
  \end{figure}
}
\fi

\def\fig10cap{
\label{fig-error_estimates}
\noteforeditor{Print this figure ONE (1) COLUMN wide.\newline}
Error analysis of the Levenberg-Marquardt method results using
the $\chi^2_\gamma$ statistic with one free parameter.
The thick curve in each panel shows the cumulative distribution
of the best-fit estimates of the slope $\gamma$.
The right (left) thin curve in each panel shows the
cumulative distribution of $\gamma$ plus (minus) $\sigma_\gamma$
which is the error estimate of the best-fit slope value.
The statistical analysis of this data is presented in
Table
\protect\ref{tbl-marquardt_1fp}.
}

\ifundefined{showfigs}{
  \figcaption{\fig10cap}
}\else{
  \clearpage
  \newpage
  \begin{figure}
  \figurenum{10}
  \vspace*{-35truemm}
  \hspace*{-0truemm}
  \epsfxsize=6.5truein
  \epsfbox{mighell.fig10.eps}
  \vspace*{-40truemm}
  \caption[]{\baselineskip 1.15em \fig10cap}
  \end{figure}
}
\fi

\def\fig11cap{
\label{fig-spectra_fits}
\noteforeditor{Print this figure ONE (1) COLUMN wide.\newline}
The simulated X-ray spectra of Fig.\ \protect\ref{fig-spectra}
now plotted with $\chi^2_\gamma$ fits.
The best fits are shown with solid curves.
The upper and lower $1$$\,\sigma$ slope estimates are shown with
long dashed curves.
}

\ifundefined{showfigs}{
  \figcaption{\fig11cap}
}\else{
  \clearpage
  \newpage
  \begin{figure}
  \figurenum{11}
  \vspace*{-35truemm}
  \hspace*{-0truemm}
  \epsfxsize=6.5truein
  \epsfbox{mighell.fig11.eps}
  \vspace*{-40truemm}
  \caption[]{\baselineskip 1.15em \fig11cap}
  \end{figure}
}
\fi

\def\fig12cap{
\label{fig-x2l}
\noteforeditor{Print this figure TWO (2) COLUMNS wide.\newline}
A comparison of the results of
the Levenberg-Marquardt method with 1 free parameter (left)
and Powell's method with 1 free parameter (right)
for three statistics:
$\chi^2_\gamma$,
$\chi^2_\lambda$,
Cash's $C$.
The statistical analysis of this data is presented in
Tables
\protect\ref{tbl-powell_1fp},
\protect\ref{tbl-powell_1fp_x2l},
and
\protect\ref{tbl-marquardt_1fp_x2l}.
}

\ifundefined{showfigs}{
  \figcaption{\fig12cap}
}\else{
  \clearpage
  \newpage
  \begin{figure}
  \figurenum{12}
  \vspace*{-20truemm}
  \hspace*{-10truemm}
  \epsfxsize=7.0truein
  \epsfbox{mighell.fig12.eps}
  \vspace*{-40truemm}
  \caption[]{\baselineskip 1.15em \fig12cap}
  \end{figure}
}
\fi

\newpage
\begin{table}
\dummytable\label{tbl-powell_2fp}
\ifundefined{showtables}{
}\else{
  \vspace*{-20truemm}
  \hspace*{-20truemm}
  \epsfxsize=8.0truein
  \epsfbox{mighell.tab01.eps}
}
\fi
\end{table}

\newpage
\begin{table}
\dummytable\label{tbl-marquardt_2fp}
\ifundefined{showtables}{
}\else{
  \vspace*{-20truemm}
  \hspace*{-20truemm}
  \epsfxsize=8.0truein
  \epsfbox{mighell.tab02.eps}
}
\fi
\end{table}

\newpage
\begin{table}
\dummytable\label{tbl-powell_1fp}
\ifundefined{showtables}{
}\else{
  \vspace*{-20truemm}
  \hspace*{-20truemm}
  \epsfxsize=8.0truein
  \epsfbox{mighell.tab03.eps}
}
\fi
\end{table}

\newpage
\begin{table}
\dummytable\label{tbl-marquardt_1fp}
\ifundefined{showtables}{
}\else{
  \vspace*{-20truemm}
  \hspace*{-20truemm}
  \epsfxsize=8.0truein
  \epsfbox{mighell.tab04.eps}
}
\fi
\end{table}

\newpage
\begin{table}
\dummytable\label{tbl-powell_1fp_x2l}
\ifundefined{showtables}{
}\else{
  \vspace*{-20truemm}
  \hspace*{-20truemm}
  \epsfxsize=8.0truein
  \epsfbox{mighell.tab05.eps}
}
\fi
\end{table}

\newpage
\begin{table}
\dummytable\label{tbl-marquardt_1fp_x2l}
\ifundefined{showtables}{
}\else{
  \vspace*{-20truemm}
  \hspace*{-20truemm}
  \epsfxsize=8.0truein
  \epsfbox{mighell.tab06.eps}
}
\fi
\end{table}

\end{document}